\documentclass[english,aps,prl,twocolumn,superscriptaddress,showpacs,amsmath]{revtex4-1}

\usepackage[T1]{fontenc}
\usepackage[latin9]{inputenc}
\usepackage{color}
\usepackage{pdfpages}
\usepackage{graphicx}
\usepackage[breaklinks=true,colorlinks=true,urlcolor=blue,
citecolor=blue,linkcolor=blue,bookmarks=false]{hyperref}

\makeatletter
\AtBeginDocument{\let\LS@rot\@undefined}
\makeatother

\begin{document}
%%%%%%%%%%%%%%%%%%%%%%%%%%%%   Title and Author
\title{Energetics of complex phase diagram in a tunable bilayer graphene probed by quantum capacitance}

\author{Manabendra Kuiri}
\author{Anindya Das}
\email{anindya@iisc.ac.in}
\affiliation{Department of Physics, Indian Institute of Science, Bangalore 560012, India}

\begin{abstract}
Bilayer graphene provides a unique platform to explore the rich physics in quantum Hall effect. The unusual combination of spin, valley and orbital degeneracy leads to interesting symmetry broken states with electric and magnetic field. Conventional transport measurements like resistance measurements have been performed to probe  the different ordered states in bilayer graphene. However, not much work has been done to directly map the energetics of those states in bilayer graphene. Here, we have carried out the magneto capacitance measurements with electric and magnetic field in a hexagonal boron nitride encapsulated dual gated bilayer graphene device. At zero magnetic field, using the quantum capacitance technique we measure the gap around the charge neutrality point as a function of perpendicular electric field and the obtained value of the gap matches well with the theory. In presence of perpendicular magnetic field, we observe Landau level crossing in our magneto-capacitance measurements with electric field. The gap closing and reopening of the lowest Landau level with electric and magnetic field shows the transition from one ordered state to another one. Further more we observe the collapsing of the Landau levels near the band edge at higher electric field ($\bar D > 0.5$ V/nm), which was predicted theoretically. The complete energetics of the Landau levels of bilayer graphene with electric and magnetic field in our experiment paves the way to unravel the nature of ground states of the system.
\end{abstract}

\maketitle
\section{Introduction}
%(\textcolor{blue}{BLG introduction :B=0?}):
Bilayer graphene (BLG) provides a unique two-dimensional system in condensed matter physics, where the low energy spectrum is gapless %and parabolic bands, with massive charge carries 
touching at K and K' points and %a potential difference applied between the two layers 
an external electric field opens up a tunable gap at the valley points\cite{PhysRevLett.96.086805,ohta2006controlling}. In clean samples the $e-e$ interactions lead to gap opening even without an external electric field\cite{PhysRevLett.104.156803,PhysRevLett.108.076602} and interesting phases like quantum spin Hall, anomalous quantum Hall\cite{PhysRevB.82.115124}, layer antiferromagnet\cite{PhysRevB.87.195413}, and nematic\cite{PhysRevB.81.041401} states were suggested to be the possible ground state at the neutrality point\cite{PhysRevLett.108.186804}. Bilayer graphene is even more interesting in presence of magnetic field due to the additional orbital degeneracy of the lowest Landau level (LL) together with spin and valley degeneracy, resulting in complex quantum Hall states (QHS)\cite{feldman2009broken,maher2014tunable}. The coupling of electric and magnetic field leads to transitions between different spin, valley and orbital ordering leading to unique interaction driven symmetry broken states\cite{PhysRevLett.120.047701,velasco2014transport,hunt2017direct,maher2013evidence,kou2014electron,lee2014chemical,PhysRevB.87.161402,velasco2012transport,PhysRevB.85.115408,PhysRevLett.107.016803,velasco2014competing}. Thus BLG provides an excellent platform to probe the phase transitions between different ordered states\cite{PhysRevB.85.235460,leroy2014emergent,PhysRevB.86.075450,PhysRevLett.109.046803}.\\

There has been extensive studies to find the nature of ordered states in BLG, both theoretically\cite{PhysRevB.86.075450,PhysRevLett.109.046803,PhysRevB.86.195435} and experimentally\cite{weitz2010broken,PhysRevLett.107.016803,lee2014chemical,hunt2017direct,zibrov2017tunable}. The model employed in Refs. \cite{PhysRevLett.109.046803,lee2014chemical,leroy2014emergent} shows that at finite magnetic field ($B$), the LLs are spin splitted and the orbital and valley degeneracies are lifted by the application of electric field. However, model employed in Refs.\cite{hunt2017direct,zibrov2017tunable} showed that at finite $B$ both the spin and orbital degeneracies are lifted, and the application of electric field results in lifting the valley degeneracy only. However, there is no common consensus about the order of the ground state of these symmetry broken states\cite{maher2013evidence,kou2014electron,hunt2017direct}.\\

Recent transport measurements in dual gated geometry have observed the crossing of LLs leading to the closing of gap which is attributed to the phase transition between different type of ordered states\cite{weitz2010broken,PhysRevLett.107.016803}. Although transport measurements can provide an indication of the gap size, but the true energetics of these states cannot be estimated by conventional transport measurements. Therefore, thermodynamic measurement is desirable to directly probe the electronic properties as well as the energetics of the these states\cite{PhysRevLett.105.256806}. The proper knowledge of the energetics of these LL crossing points together with the variation of LL energy by external electric and magnetic fields provide key insights to the nature of ground state, which has been employed to probe the magnetization of quantum hall states\cite{de2000resistance} and many body enhanced susceptibility\cite{PhysRevLett.90.056805} in two dimensional electron gas (2DEG).\\

In order to obtain the energetics in BLG with electric and magnetic field, we employ magneto capacitance studies in a hexagonal boron nitride (hBN) encapsulated dual gated BLG device. At zero magnetic field, using our quantum capacitance measurement we measure the gap around the charge neutrality point as a function of perpendicular electric field($\bar D$), where the obtained value of the gap matches well with the previously reported values\cite{zhang2009direct}. In presence of perpendicular magnetic field, we observe LL crossing in our magneto-capacitance measurements with $\bar D$. The gap closing and reopening of the lowest LL with $\bar D$ and $B$ shows the transition from one ordered state to another one. The values of critical electric field ($\bar D_c$) required to close the gap as a function of magnetic field matches well with the earlier reports\cite{weitz2010broken,PhysRevLett.107.016803}. We further obtain the energetics of the LLs as a function of $\bar D$ and $B$, where the renormalization of  LL spectrum at higher electric field ($\bar D > 0.5$ V/nm) is clearly visible. %We further study the effect of LL spectrum renormalization, by breaking the inversion symmetry at high electric fields.
 It has been shown theoretically that at higher electric fields the LLs collapses at the band edge due to LL coupling and hybridization \cite{PhysRevLett.96.086805,PhysRevB.73.245426,PhysRevB.87.075417}, which has not been observed experimentally prior to this report. %Our magneto-capacitance measurement at higher $\bar D$ ($\bar D > 0.5$ V/nm) captures the above mentioned theoretical predictions.

%As described thereotically in Ref\cite{PhysRevLett.96.086805,PhysRevB.73.245426}, introducing electric field between the layers leads to selective control of charge carriers in each layer, which may result in lifting the intervalley degeneracy leading to LL coupling and hybridization\cite{PhysRevB.87.075417}. This leads to LL crossing and anticrossing\cite{PhysRevB.87.075417} for sufficiently high $\bar D$. Thus BLG system can be promising candidate to study quantum hall ferromagnetism[ ]. However, the energetics of the LL spectrum with high broken inversion symmetry (large $\bar D$) has not been experimentally studied yet.
%Using dual gated geometry the independent control over carrier density ($n$) and electric field ($D$), allows one to access experimentally different complex phases in bilayer graphene with magnetic field. \\

%Capacitance spectroscopy provides an alternate powerful tool to probe the low dimensional system revealing electron-electron interactions\cite{PhysRevLett.68.674,PhysRevB.50.1760,ilani2006measurement}, many body effects, layer polarization\cite{PhysRevB.85.235458,hunt2017direct}, Fermi velocity renormalization \cite{yu2013interaction}, band structure modification\cite{PhysRevB.92.075408,wang2013negative}. Capacitance studies are particularly important for disordered systems, to probe localized states which are usually suppressed in conventional transport measurements\cite{kuiri2015probing}.

%%%%%%%%%%%%%%%%%%%%% Figure 1 %%%%%%%%%%%%%%%%%%%%%%%%%%%%%%%%%%%%
\begin{figure*}[ht!]
 \includegraphics[width=0.8\textwidth]{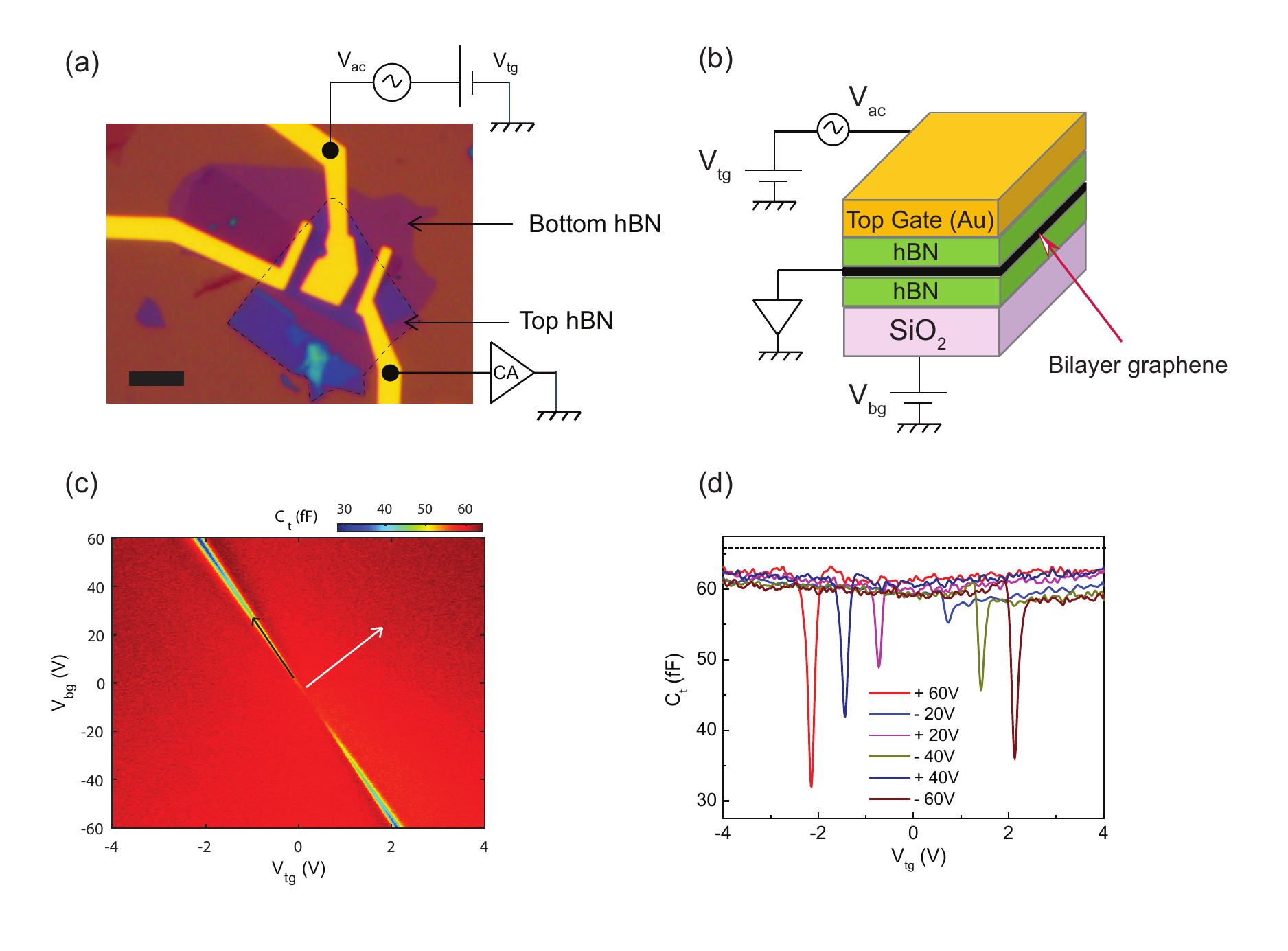}
 \caption{(Color Online) (a) Optical image of the device. Scale bar $5\mu m$. (b) Schematic of the device architecture and measurement scheme. (c) Color plot of measured total capacitance($C_t$) as function of backgate voltage ($V_{bg}$) and topgate voltage ($V_{tg}$) at T$\sim$ 240 mK. Black solid line shows the $\bar D$ axis, and white dashed line shows the $n$ axis. (d) Cut lines showing $C_t$ as a function of topgate voltage($V_{tg}$) for several values of backgate voltage ($V_{bg}$).}
 \label{fig:fig1}
\end{figure*}
%%%%%%%%%%%%%%%%%%%%%%%%%%%%%%%%%%%%%%%%%%%%%%%%%%%%%%%%%%%%%%%%%%%%%%%%%%

\section{experimental details}
Dual gated bilayer graphene device was fabricated using van der Waals assembly, following the procedure developed by Wang \textit{et.al. }\cite{wang2013one}. Briefly bilayer graphene (BLG) was first mechanically exfoliated onto a piranha cleaned Si/SiO$_2$ substrate from bulk single crystal of natural graphite. On another clean substrate hBN was mechanically exfoliated and potential thin hBN was looked for using optical microscope. Using dark field microscope imaging hBN flake with uniform smooth surface and free of bubbles was chosen. hBN, BLG and hBN were picked up sequentially one on top of another and the complete stack (hBN-BLG-hBN) was deposited onto a $n++$ doped Si/SiO$_2$ substrate with 285 nm oxide. The stack was then annealed at 200$^\circ$C in vacuum to get a uniform surface free of bubbles. The electrical contacts were fabricated using electron beam lithography followed by etching the hBN-BLG-hBN stack, and one-dimensional contact was established by thermally evaporating Cr/Au (5nm/70nm)\cite{wang2013one}. Another step of lithography and thermal deposition was carried out to define the topgate electrode (see supplemental material; SM-Sec.I for details). The optical image of the final device is shown in Fig.~1a. The schematic of the device and the measurement scheme are shown in Fig.~1b. The top hBN thickness $\sim$ 11 nm and bottom hBN thickness $\sim$ 15 nm were measured using atomic force microscope (see SM-Sec.II). The thickness of top hBN was found independently using period of oscillation of the capacitance minima in magnetic field \cite{yu2013interaction}. The excellent dielectric properties of hBN serves the purpose of using thin gate dielectric for measuring detectable change in total capacitance (C$_t$). All the measurements were carried out in a $^3$He refrigerator with a base temperature $T\sim $ 240 mK.\\

For the capacitance measurements we have used the measurement scheme described in our earlier works\cite{bppaper,PhysRevB.98.035418} using a home built differential current amplifier with a gain of $10^7$. The capacitance has been measured between the topgate electrode and BLG with a small ac excitation voltage of $\sim$ 10-15 mV at a frequency of $\sim$ 5 kHz with a resolution of $\sim 0.5~fF$. All wires were shielded to reduce the parasitic capacitance. In a parallel plate capacitor made of a normal bulk metal and a two dimensional material like graphene, adding a charge requires  electrostatic energy, but also kinetic energy due to the change in chemical potential, thereby contributing to the total capacitance\cite{luryi1988quantum}. The total measured differential capacitance in such a system is given by 

\begin{equation}
C_t= \left(\frac{1}{C_g} + \frac{1}{C_q} \right)^{-1} +C_p   
\label{eqn:eq1}
\end{equation}
where, $C_g$ is the geometric capacitance, $C_q=Se^2\frac{dn}{d\mu}$ is the quantum capacitance; $e$ is the electronic charge; $S$ is the area under the topgate electrode; $\frac{dn}{d\mu}$ is the thermodynamic compressibility, $C_p$ is the parasitic capacitance arising due to the wirings plus the stray capacitances. In BLG, the application of electric field between the layers results in breaking the inversion symmetry, which in turn opens up a band gap \cite{zhang2009direct} at the charge neutrality point. Dual gated geometry allows us to independently control electronic density ($n$) and electric displacement field ($\bar D$) under the topgated region. The net transverse electric field in a dual gated device is given by $\bar D=[C_{bg}(V_{bg}-V_{bg}^0)-C_{tg}(V_{tg}-V_{tg}^0)]/2\epsilon_0$ and the total carrier density is given by $n=[C_{bg}(V_{bg}-V_{bg}^0)+C_{tg}(V_{tg}-V_{tg}^0)]/e$ ; $\epsilon_0$ is the vacuum permittivity, $e$ is the electronic charge,  $C_{bg}(C_{tg})$ is the capacitance per unit area of the backgate(topgate) region and $V_{bg}^0,V_{tg}^0$ are the charge neutrality points.

%%%%%%%%%%%%%%%%%%%%% Figure 3%%%%%%%%%%%%%%%%%%%%%%%%%%%%%%%%%%%%
\begin{figure*}[ht!]
 \includegraphics[width=1\textwidth]{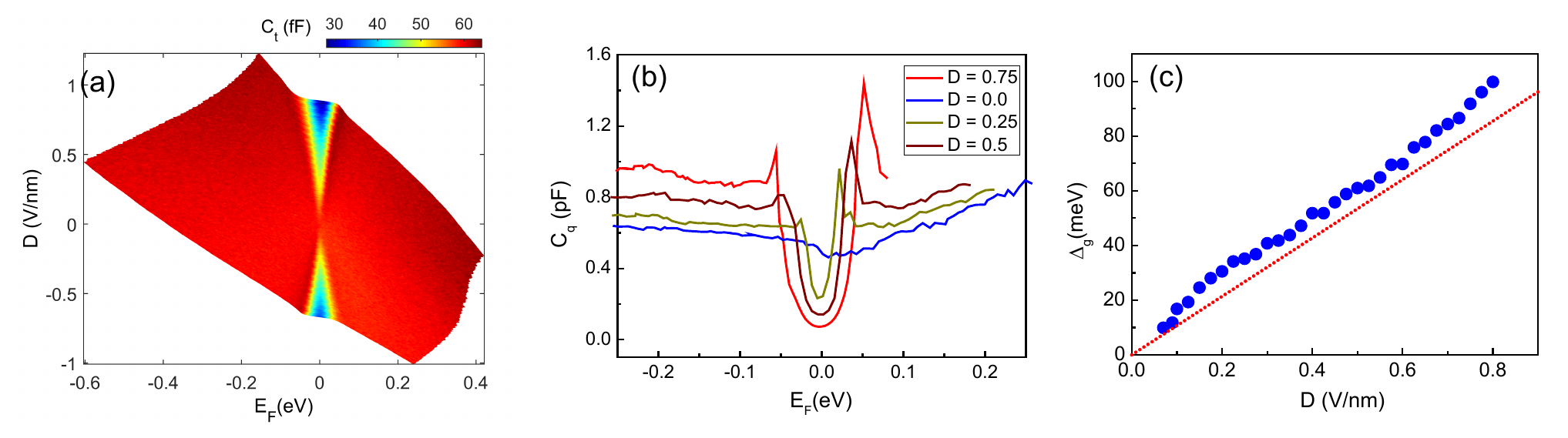}
 \caption{(Color Online) (a) Colorplot of total capacitance ($C_t$) as a function of electric field ($\bar D$) and Fermi energy $E_F$. (b) Extracted quantum capacitance ($C_q$) with $E_F$ for different value of $\bar D$. (d) Blue scattered plots shows the extracted band gap as a function of electric field $\bar D$. Red dashed line shows the calculated band gap with $\bar D$ (Refs. \cite{PhysRevB.75.155115,lee2014chemical}).}
 \label{fig:fig2}
\end{figure*}
%%%%%%%%%%%%%%%%%%%%%%%%%%%%%%%%%%%%%%%%%%%%%%%%%%%%%%%%%%%%%%%%%%%%%%%%%%

\section{Capacitance Data at B=0T}
%To show the efficacy of our quantum capacitance measurement we show capacitance data for $B=0$. 
Fig.~1c shows the colorplot of the measured total capacitance, $C_t$ as a function of backgate voltage ($V_{bg}$) and topgate voltage ($V_{tg}$) at B = 0T. The data was taken by sweeping the topgate voltage for different values of backgate voltages. Tuning of topgate and backgate changes both the total carrier density ($n$) and the band gap ($\Delta_g$). The diagonal white dashed marked in Fig.~1c shows the direction of $n$ and solid black line shows the direction of $\bar D$. For $\bar D\sim 0$, C$_t$ exhibits a minimum at zero density, signifying the hyperbolic nature of band structure for ungapped bilayer graphene\cite{PhysRevB.82.041412}. As $|\bar D|$ increases the capacitance minima decreases revealing the formation of gap in the energy spectrum in bilayer graphene\cite{PhysRevB.85.235458}. The diagonal line in Fig.~1c corresponds to the charge neutrality point under the topgated region. Along the diagonal line the capacitance minima decreases signifying the electric field induced band gap opening. The charge neutrality points ($V_{bg}^0,V_{tg}^0$) are located at 0.3V, -8.5V. From the slope of the diagonal line we can effectively estimate the ratio of the capacitive coupling between the top and bottom gates $C_{tg}/C_{bg} \sim 27$ ($C_{tg}\Delta V_{tg}=C_{bg}\Delta V_{bg}$ along the diagonal line in Fig.~1c, $d_{bg}\sim$ 300 nm, $\epsilon_{hBN}=\epsilon_{SiO_2}\sim~3.9$, yields $d_{tg}\sim$ 10.75 nm, which matches well with the value of $d_{tg}\sim$ 11 nm obtained using AFM, see SM-Sec.II). Fig.~1d shows the cut lines of $C_t$ as a function of $V_{tg}$ for several value of $V_{bg}$. The geometric capacitance $C_g\sim 66~fF$ is marked with dashed black line. Noting the area of our device $S\sim 21~\mu m^2$, the effective geometric capacitance was $C_g\sim 66~fF$. The parasitic capacitance was estimated by comparing the experimental capacitance data at $\bar D=0$ with the theoretical one (Eq.\ref{eqn:eq1}), where only adjusting parameter was C$_p$ (see SM-Sec.III; the density of states for ungapped bilayer graphene with effective mass $m_*=0.03m_e$ was calculated from Ref\cite{PhysRevLett.96.086805}).%fitting Eq. (\ref{eqn:eq1}), assuming the DOS for ungapped ($\bar D=0$) bilayer graphene\cite{PhysRevLett.96.086805}, with effective mass $m_*=0.03m_e$, $C_g=66fF$, with only one fitting parameter C$_p$ (see SM, Sec.3).
 The parasitic capacitance C$_p$ in our device is $\sim$ 152 fF. This value of C$_p$ is subtracted from all the data presented in this paper. \\
%%%%%%%%%%%%%%%%%%%%% Figure 4%%%%%%%%%%%%%%%%%%%%%%%%%%%%%%%%%%%%
\begin{figure*}[ht!]
 \includegraphics[width=1\textwidth]{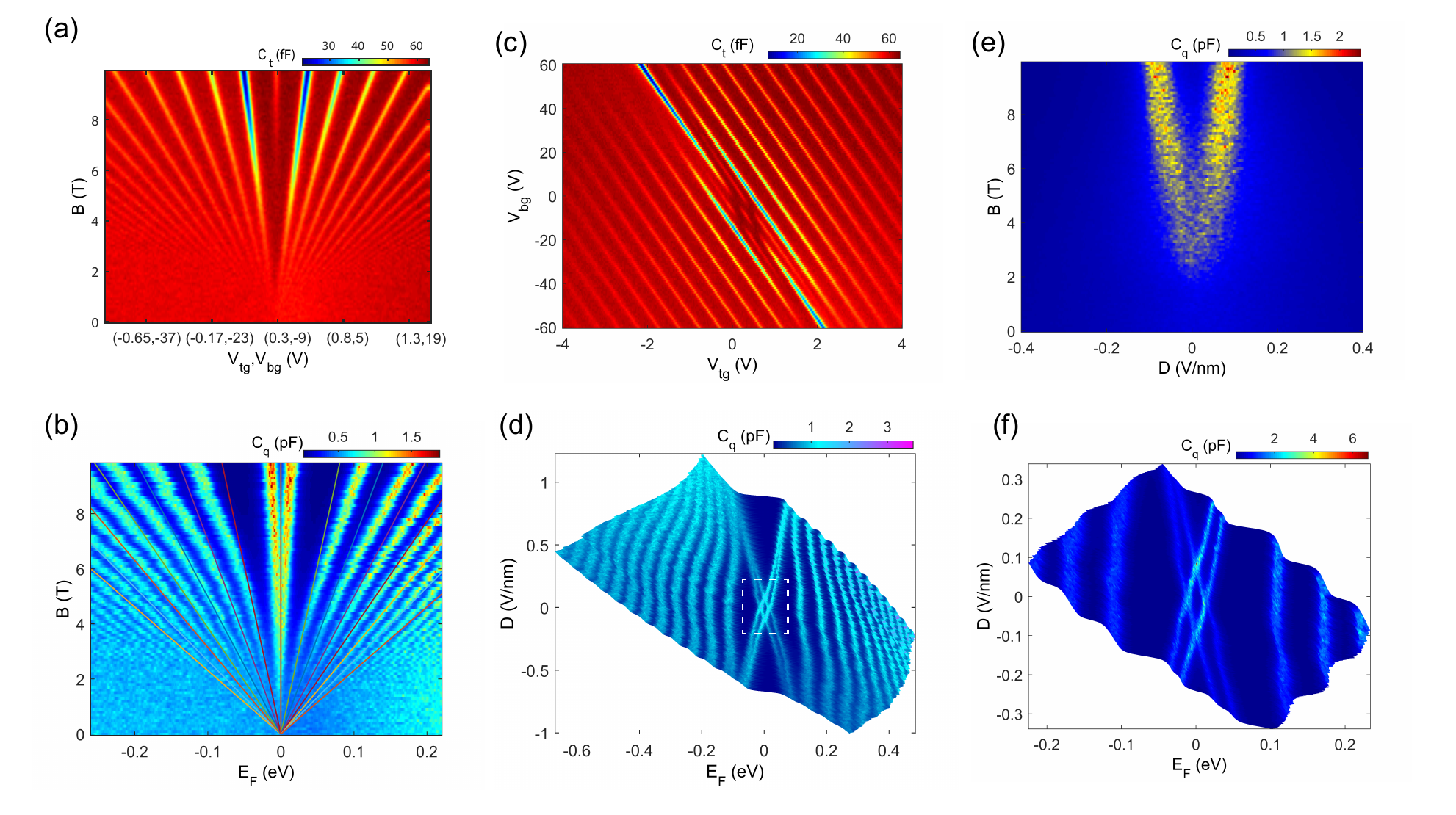}
 \caption{(Color Online) (a) Colorplot of the measured total capacitance ($C_t$) for $\bar D=0$ as a function of magnetic field ($B$). The data was recorded synchronously by sweeping $V_{tg}$ ($V_{tg}$) keeping $\bar D=0$. (b) Extracted quantum capacitance ($C_q$) as a function of $E_F$ for different values of magnetic field for $\bar D=0$. The solid lines are the single particle LL energy spectrum of BLG as discussed in the main text. (c) Colorplot of the measured total capacitance ($C_t$) as a function of ($V_{tg}$, $V_{bg}$) for $B=10~T$. (d) Extracted quantum capacitance ($C_q$) as a function of Fermi energy $E_F$ and electric field ($\bar D$) at $B=10~T$. (e) Colorplot of $C_q$ as a function of $\bar D$, and $B$ for $E_F=0$. (f) $C_q$ as a function of $E_F$, for small value of $\bar D$ (zoomed region of Fig. 5d labelled in white dashed box).}
 \label{fig:fig3}
\end{figure*}
%%%%%%%%%%%%%%%%%%%%%%%%%%%%%%%%%%%%%%%%%%%%%%%%%%%%%%%%%%%%%%%%%%%%%%%%%%

In order to get a better insight to the experimental data we need to extract the quantum capacitance ($C_q$) as a function of Fermi energy ($E_F$) from the experimentally measured $C_t$ as a function of backgate and topgate voltages . The Fermi energy and band gap are independently controlled by changing $V_{bg}$ and $V_{tg}$. Thus the quantum capacitance should be extracted along the constant $\bar{D}$ lines as a function of Fermi energy. We have followed a similar approach as described in Ref \cite{kanayama2015gap} (see SM-Sec.IV for details). %It must be noted that the contribution of $C_{bg}$ is embedded through $C_q$ and $E_F$\cite{kanayama2015gap}.
 The Fermi energy of bilayer graphene is given by the charge conservation relation $E_F=e\int_{0}^{V_{tg}}\left(1-\frac{C_t}{C_g}\right)dV_{tg}$ \cite{droscher2010quantum}. Fig.~2a shows the colorplot of total capacitance ($C_t$) as a function of Fermi energy and electric field. It can be seen that the band gap opens with the increment of $\bar D$. The maximum $\bar D$ we could reach was 0.8 V/nm with a band gap opening $\Delta_g \sim$ 80 meV in agreement with previously reported values\cite{zhang2009direct}. The extracted quantum capacitance ($C_q^{-1}=C_t^{-1}-C_g^{-1}$) for several values of $\bar D$ is shown in Fig.~2b. It can be seen that with the increment of $\bar D$, $C_q$ decreases signifying the increase of band gap. We have observed asymmetry in the $C_q$ for the electron and hole side, which has also been previously observed by other groups\cite{PhysRevB.85.235458,kanayama2015gap}. The 1/$\sqrt{E}$ van hove singularity is also observed at the band edge as predicted\cite{mccann2013electronic}. The extracted $\Delta_g$ as a function of $\bar D$ has been shown in Fig. 2c. The measured band gap values matches well with the theoretical band gap calculated using tight binding model\cite{PhysRevB.75.155115}.

\section{Magneto-capacitance data}
The competing magnetic and electric field leads to various interesting phases in the LL spectrum of BLG. To visualize the energetics of the LLs as a function of $\bar D$ and B, we present our magneto capacitance data. %Conventional transport measurements probe the chiral edge state of BLG in the quantum hall effect (QHE). In contrast quantum capacitance probes the average DOS of the bulk in the QHE revealing better insights to the LL spectrum.
 For an ungapped pristine BLG, in absence of any interactions, the LL energies in a perpendicular magnetic field is given by $E_N=\pm\hbar\omega_c \sqrt{N(N-1)}$, where $\omega_c=eB/m^*$ is the cyclotron frequency, and $N=0,\pm1,\pm2...$ are the orbital index. For $N=0,1; ~E_N=0$. Thus, the zeroth energy LL is eight-fold degenerate, whereas all other landau levels ($N\geq2$) are four fold degenerate (two spin and two valley)\cite{PhysRevLett.96.086805}.\\

%%%%%%%%%%%%%%%%%%%%% Figure 5%%%%%%%%%%%%%%%%%%%%%%%%%%%%%%%%%%%%
\begin{figure}[t!]
 \includegraphics[width=0.35\textwidth]{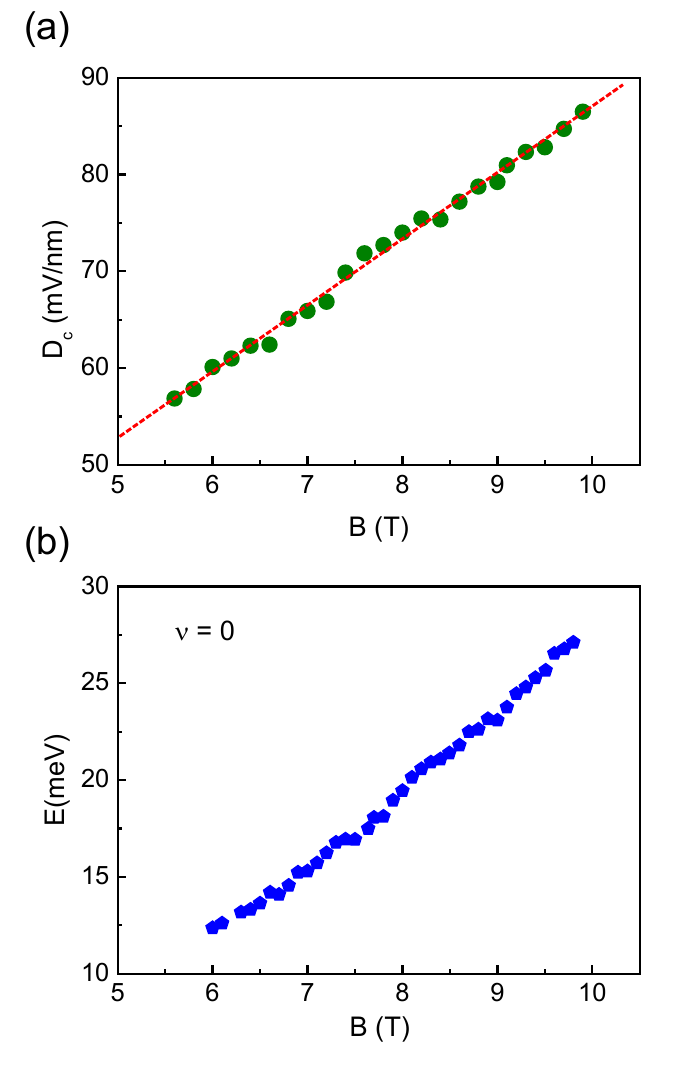}
 \caption{(Color Online) (a) Critical electric field ($\bar D_c$) as a function of $B$. (b) The LL energies for $\nu=0$ state as a function of $B$ for $\bar D=0$. }
 \label{fig:fig4}
\end{figure}
%%%%%%%%%%%%%%%%%%%%%%%%%%%%%%%%%%%%%%%%%%%%%%%%%%%%%%%%%%%%%%%%%%%%%%%%%%
%%%%%%%%%%%%%%%%%%%% Figure 6%%%%%%%%%%%%%%%%%%%%%%%%%%%%%%%%%%%%
\begin{figure*}[ht!]
 \includegraphics[width=0.8\textwidth]{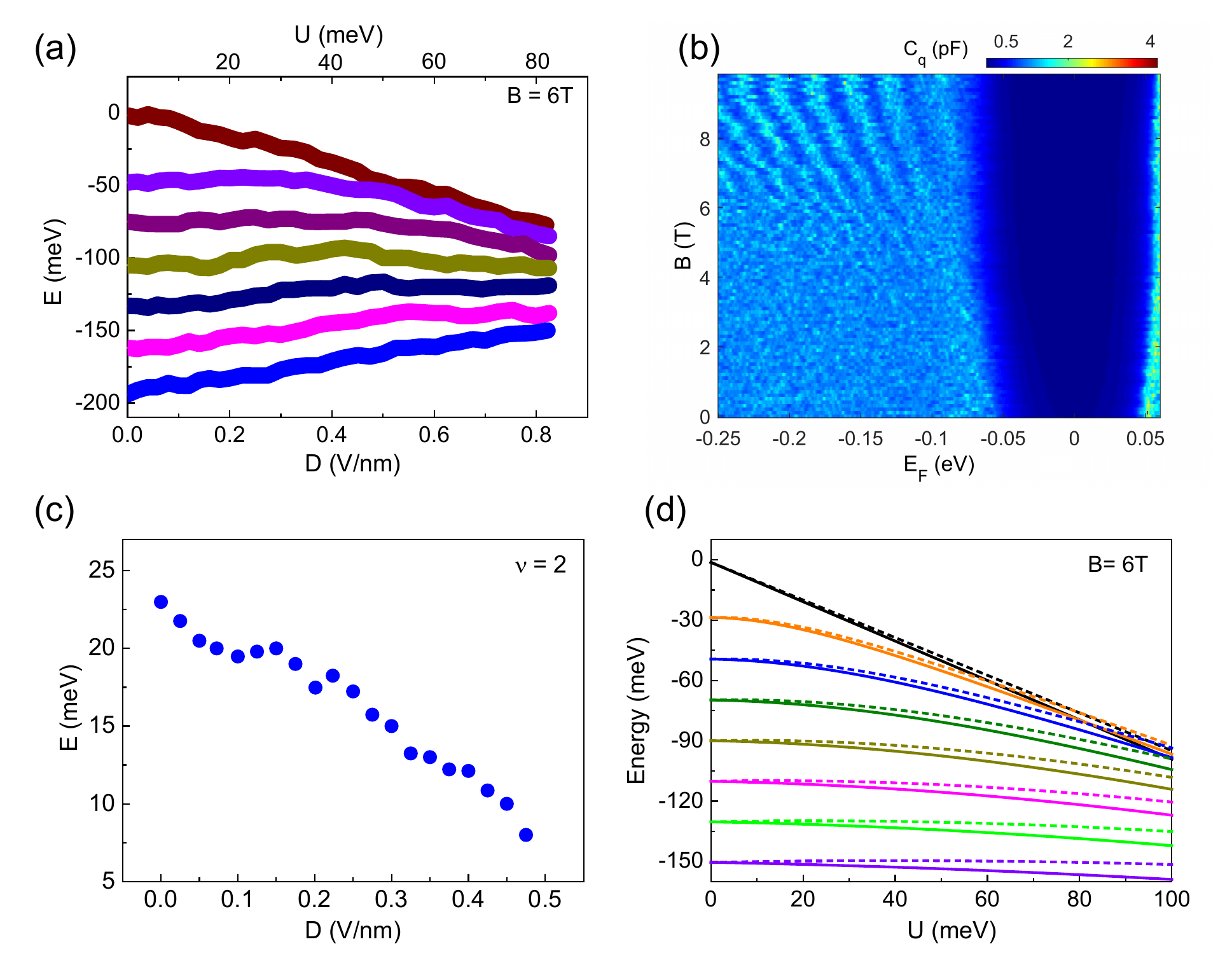}
 \caption{(Color Online) (a) LL energies as a function of electric field ($\bar D$) for B=6T. (b) Extracted quantum capacitance ($C_q$)as a function of $E_F$ for different values of magnetic field at $\bar D=0.8~V/nm$. (c) Evolution of the $\nu=2$ gap as a function of electric field for $B=10T$. (d) Theoretical LL energies as a function of interlayer bias (U) for B=6T.}
 \label{fig:fig5}
\end{figure*}
%%%%%%%%%%%%%%%%%%%%%%%%%%%%%%%%%%%%%%%%%%%%%%%%%%%%%%%%%%%%%%%%%%%%%%%%%%

Fig.~3a shows the experimental LL fan diagram for $\bar D=0$. Here, $C_t$ was measured by sweeping $V_{tg}$,$V_{bg}$ synchronously keeping the $\bar D=0$ and changing only the carrier density. The dips in the capacitance data corresponds to the LL gap. The gap around the zeroth LL start appearing for $B>5T$. The LL corresponding to $N=\pm2, \pm3,\pm4$ can be seen in Fig. 3a. The geometric capacitance C$_g$ was determined independently from the fact that spacing $\Delta V_{tg}$ between the adjacent capacitance minima in Fig.~3a is given by the amount of charge required to fill each Landau level\cite{yu2013interaction} ($C_g\Delta V_g = \frac{4Se^2B}{h}$, where $\Delta V_g\sim 0.48V$) (see SM-Sec.V) yielding an effective $C_g\sim65.5~fF$ which matches quite well as extracted from the colorplot of Fig. 1c and AFM imaging (see SM-Sec.II). The conversion of $x-$ axis in Fig. 3a, which is a combination of topgate voltage and backgate voltage, to Fermi energy is shown in SM-Sec.VI. Fig.~3b shows the result of such a conversion where we plot the extracted $C_q$ as a function Fermi energy for different values of magnetic field. The solid lines are generated using single particle LL energies for ungapped BLG ($E_N=\pm\hbar\omega_c \sqrt{N(N-1)}$, with effective mass $m_*=0.03m_e$). It can be seen that upto $B<6T$, the extracted LL spectrum matches quite well with the theory. However, for $B>6T$ we observe noticeable mismatch between the experimental and the theoretical values (10\%-20\%), which has also been addressed in previous studies, employing magneto-capacitance measurements\cite{yu2013interaction,yu2014hierarchy,PhysRevB.98.035418}. This mismatch has been attributed to the inaccurate conversion in determining $E_F$ at higher magnetic field as the bulk becomes more insulating and increasingly isolated from electrical contacts leading to excess deep in the $C_t$ versus gate voltage curve.\\

We now show the LL spectrum as a function of electric and magnetic field. Figure.~3c shows the measured $C_t$ as a function of $V_{bg}$ and $V_{tg}$ for $B=10T$. In Fig.~3d we have shown the extracted quantum capacitance as a function of $E_F$ and $\bar D$ for B=10T. The parallel lines are the different LLs which evolves with $\bar D$. The most striking feature is the evolution of the zeroth energy LL with $\bar D$. In Fig.~3f we have shown the zoomed part of the Fig. 3d (white dashed box). %to note is that the gap opening and gap closing of the lowest LL ($N=0,1$), which is a signature of phase transition\cite{leroy2014emergent}.
 The emergence of the $\nu=0$ insulating state can be seen for $\bar D=0$. With the increment of $\bar D$, we see the evolution of the $\nu=0$ insulating state. For small values of $\bar D$, $\nu=0$ state remains gapped, with increase in $\bar D$, the gap decreases monotonically, and then for a critical value $\bar D_c=0.08$mV/nm the gap closes, further increase in $\bar D$ the gap again re-opens and remains gapped for high $\bar D$ (maximum $\bar D$ for our device was $\bar D\sim $1V/nm). This electric field induced gap closing and re opening is a signature of phase transition\cite{leroy2014emergent}. In Fig.~3e we show the evolution of the $\nu=0$ state with $\bar D$ and $B$. Here, the topgate and backgate were swept synchronously to maintain zero carrier density and vary only $\bar D$ as described earlier. The $V$ shaped yellow structure in Fig. 3e separate out two insulating states (blue regions inside and outside of the $V$), which is in consistent with earlier reports.
Fig.~4a shows the plot of critical electric field, $\bar D_c$ as a function of B. The $\bar D_c$, which determines the transition point, can be written as a linear function of magnetic field as $\bar D_c=\bar D_{off}+\alpha B$, where $\bar D_{off}$ is the offset electric field and $\alpha$ is the slope. For our case $\bar D_{off}=18$ and $\alpha=7~mV/nm~\times B[T]$ matches well with the theoretically predicted values\cite{PhysRevB.83.115455} and experimentally observed values for $\bar D_c$ reported using resistance measurements\cite{PhysRevLett.107.016803}, where the $\nu=0$ QHS undergoes a phase transition between the spin polarized phase and the layer polarized phase in the ($B-\bar D$) plane. Further more the $\nu=0$ gap at $\bar D = 0$ as a function of B is shown in Fig.~4b, where the gap increase linearly with $B$, with a slope of $3~meV/T$, in agreement with previous reports\cite{PhysRevLett.105.256806}, which suggests the ground state is spin polarized ($\bar D = 0$) and rules out the possibility that the ground state is valley polarized\cite{kou2014electron}. %With the above presented data, we conclude that the ground state in our system is Spin polarized in consistent with earlier reports in BLG.

%%%%%%%%%%%%%%%%%%%%% Figure 6%%%%%%%%%%%%%%%%%%%%%%%%%%%%%%%%%%%%
%\begin{figure}[ht!]
% \includegraphics[width=0.3\textwidth]{fig7.pdf}
% \caption{(Color Online) (a) $\nu=2$ gap for $B=10~T$ as a function of $\bar D$}
% \label{fig:fig5}
%\end{figure}
%%%%%%%%%%%%%%%%%%%%%%%%%%%%%%%%%%%%%%%%%%%%%%%%%%%%%%%%%%%%%%%%%%%%%%%%%%%
\section{Landau Levels with high D}
Theoretical work employing tight binding calculations have shown that the existence of interlayer bias between the layers ($U$) will have compelling effect on the LL spectrum of BLG\cite{PhysRevB.76.115419}. In this section we will discuss about the evolution of LL spectrum with high interlayer bias. Figure 5a shows the LL energies as a function of $\bar D$ for B=6T. One striking feature  is the reduction of the energy separation between the LLs as the band gap increases, specially between the LLs near the band edge. %We observed that with increase in $\bar D$, LL tends to bend and then start collapsing at high $\bar D$.
 For $\bar D > 0.5~ V/nm$ we see the LLs near the band edge merge with each other. Fig.~5b shows colorplot of LL spectrum (as $C_q$) as a function of $E_F$ for $\bar D=0.8~V/nm$ ($\Delta_g\sim 80~meV$). One can clearly see the differences between the LL spectrum at $\bar D=0$ (Fig. 3b) and $\bar D=0.8~V/nm$ (Fig. 5b). At $\bar D=0$ the LLs are clearly visible at B = 2T where as at $\bar D=0.8~V/nm$ LLs can be hardly seen even at B = 8T. It can be also seen from the Fig. 5b that the LLs are broadened and the broadening is higher for lower LLs near the band edge. In Fig. 5c, we also show the evolution of the gap for $\nu=2$ state as a function of $\bar D$ for B = 10T. One can notice that for a fixed magnetic field the LLs gap decreases almost linearly with increasing $\bar D$. It has been shown theoretically in Ref\cite{PhysRevB.84.075451} that the LL spectrum in presence of B and $\bar D$ has the following energy eigenvalues for $n>0$

\begin{equation}
\begin{split}
E^{\pm}_{n,s_1,-}&=\bigg(n+\frac{1}{2}\bigg)\beta\widetilde\Delta\mp \beta U\\
&+ s_1 \sqrt{ \Big[ (2n + 1)\beta U \mp \dfrac{\beta \widetilde\Delta}{2} - U \Big]^2 \!\! + n (n + 1)\,\beta^2 \gamma_1^2\,} \,.
\end{split}
\label{eqn:E_low_energy}
\end{equation}
where, $\gamma_1=0.4~eV$, $\widetilde\Delta=59~meV$, $\beta =\frac{\omega_0^2}{\gamma_1^2}, \omega_0=\sqrt{2}\frac{\hbar v_0}{l_B}$; $l_B$ is the magnetic length, and $v_0 = \sqrt{3} \gamma_0 a_0/\hbar\approx 1.0\times10^8\,{\rm cm}/{\rm s}$ is the Fermi velocity. Fig.~5d shows the calculated LL energies as a function of energy gap (U) for B=6T. The solid and the dashed lines correspond to $K$ and $K'$ valleys. The LLs start to merge for $U>50~meV$ which matches well with the experimentally observed values as can be seen in Fig. 5a. We do not observe the splitting of the K and K' valleys due to the large broadening of our device ($\delta E_F\sim 20$ meV). Instead we observe the broadening of the LLs with increasing $\bar D$. \\

\section{conclusion}
In summary, we have mapped the complete energetics of the Landau level spectrum in a bilayer graphene with magnetic and electric field. We model a possible ground state based on our observations. We have also demonstrated the smearing of the LLs at high broken inversion symmetry in agreement with theoretical predictions.

\bibliography{ref_blgAUG}{}
\onecolumngrid
\newpage
\thispagestyle{empty}
\mbox{}


\section*{Supplementary material}

# Supplemental Material

Manabendra Kuiri[1] and Anindya Das[1]

[1]*Department of Physics, Indian Institute of Science, Bangalore 560012, India*

## I. DEVICE FABRICATION

Hexagonal boron nitride (hBN) encapsulated bilayer graphene (BLG) device was fabricated using the dry pick up technique [1]. hBN/BLG/hBN heterostructure was etched using $CHF3+O_2$ plasma to expose the graphene edge[1]. Cr/Au (5nm/70nm) was thermally deposited at a base pressure of $3\times10^{-7}$mbar to establish the edge contacts in our device. Another step of ebeam lithography was carried out followed by thermal deposition of Cr/Au (5nm/70nm) to define the topgate electrode. Figure S1 shows the optical image of the stacking process.

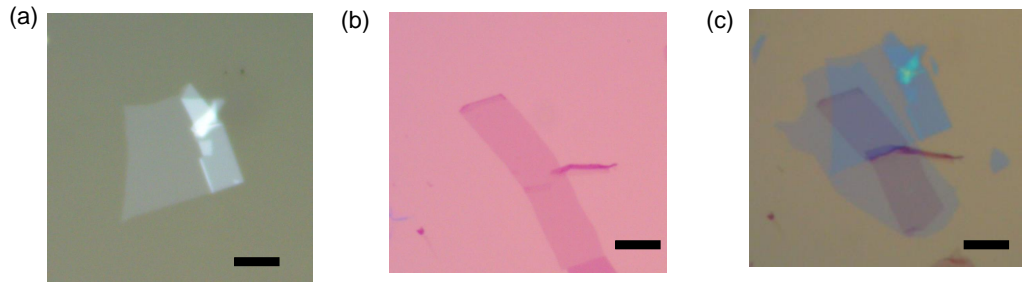

Fig. S1: (Color Online) (a) Optical image of top hBN on PPC. (b) Optical image of BLG on Si/SiO$_2$. (c) hBN/BLG/hBN stack on Si/SiO$_2$. Scale bar 4 $\mu$m.

## II. DEVICE CHARACTERIZATION

Figure S2a shows the optical image of the device. The AFM image of the top and bottom hBN is shown in Fig. S2b. The height profile for top (bottom) hBN is shown in Fig. S2c and S2d. The thickness of bottom (top) hBN are $\sim 15$ (11)nm respectively.

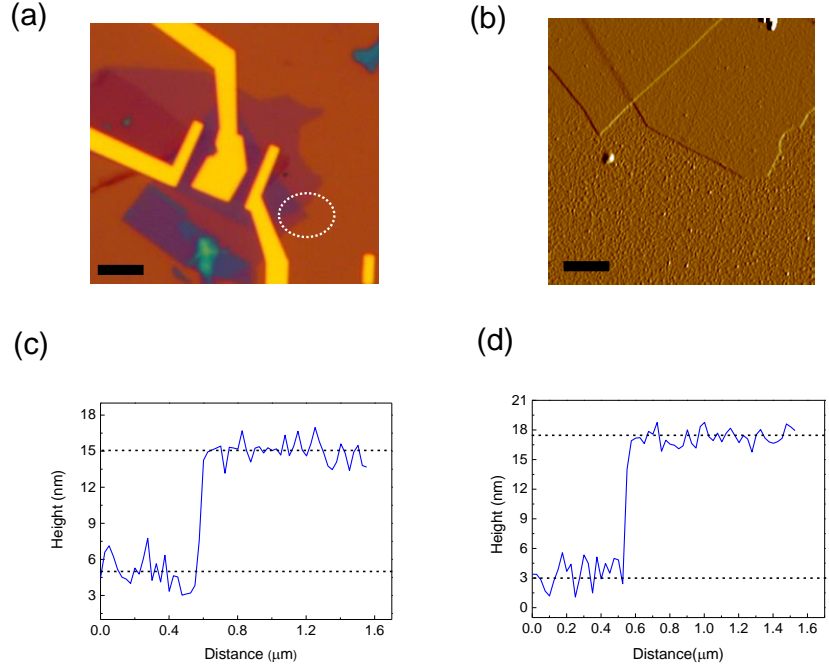

Fig. S2: (Color Online) (a) Optical image of the device. Scale bar 4 $\mu$m. (b) AFM image of the device in the region marked by white circle in (a). Scale bar 200 nm. (c) and (d) Height profile of top and bottom hBN.

## III. ESTIMATION OF PARASITIC CAPACITANCE

The total measured capacitance between the top gate and BLG is given by

$$C_t = \left( \frac{1}{C_g} + \frac{1}{C_q} \right)^{-1} + C_p \tag{1}$$

where, $C_g \sim 66 fF$ is the geometric capacitance, $C_q = Ae^2 \frac{dn}{d\mu}$ is the quantum capacitance. The area of the topgate electrode $A \sim 21 \ \mu m^2$ was measured using the optical microscope image. Using the theoretical low energy DOS of bilayer graphene as described in Ref.[2], we plot the theoretical $C_t$ as a function of carrier density $(n)$ in Fig. S3. Experimentally we measure $C_t$ as a function of backgate and topgate voltages along the $\bar{D} = 0$ line. The total carrier density $(n)$ was calculated using $n = [C_{bg}(V_{bg} - V_{bg}^0) + C_{tg}(V_{tg} - V_{tg}^0)]/e$, where $(V_{bg}^0, V_{tg}^0)$ corresponds to charge neutrality point, $C_{tg} \sim 0.0032 \ F/m^2$ and $C_{bg} \sim 1.15 \times 10^{-4} \ F/m^2$. By comparing the experimental and the theoretical plot in Fig. S3, we estimate the parasitic capacitance $\sim 152$ fF.

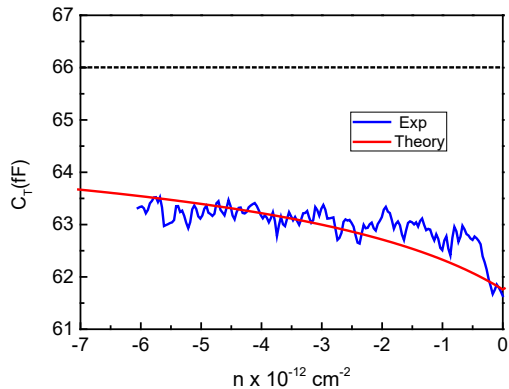

Fig. S3: (Color Online) Total capacitance ($C_t$) as a function of carrier density ($n$) experimental (blue) and theoretical (red). $C_g$ is marked with black dashed line.

## IV.  CONVERSION OF EXPERIMENTAL DATA $C_q$ AND $E_F$

As described in the main text, changing backgate voltage($V_{bg}$) and topgate voltage($V_{tg}$) changes both the Fermi energy ($E_F$) and the bang gap ($\Delta_g$) in bilayer graphene. Therefore, integrating $C_t$ as a function $V_{tg}$ for a constant $V_{bg}$ will give overestimated values of $E_F$. The correct way is to extract $E_F$ along the constant $\bar{D}$ line.

The complete capacitive circuit of the measured dual gated BLG device is shown in Fig. S4a, where $C_{tg}$ is the topgate capacitance, $C_{bg}$ is the backgate capacitance. The simplified equivalent circuit is shown in Fig. S4b, where the contribution of $C_{bg}$ in $C_t$ is embedded through $C_q$ and $E_F (= eV_{ch})$ [2]. $V_{ch}$ is the channel potential and can be extracted by

$$V_{ch} = \int_0^{V_{tg}} \left(1 - \frac{C_t}{C_g}\right) dV_{tg} \qquad (2)$$

for a constant $V_{bg}$, where $C_g \sim 66fF$ is the geometric capacitance. Figure S4c shows the illustration of the conversion procedure. The blue solid line shows the $\bar{D} = 0$ line. Now in order to convert ($V_{bg}, V_{bg}$) to Fermi energy, the integration must be carried out along $\bar{D} = 0$ line (i.e along $S - a0$, all the way to $S - a5$, along $S - X$ line) marked in Fig. S4c. This is analogous to integration of $a0 - 0$, $a1 - 1$ and iterating all the values for constant bacgkate voltages along the $S - X$ lines. This gives the Fermi energy along the $\bar{D} = 0$. Similar procedure can be followed for other constant $\bar{D}$, yielding a colorplot as shown in Fig 2a in the main manuscript. Similar converstion procedure was followed in Ref [2].

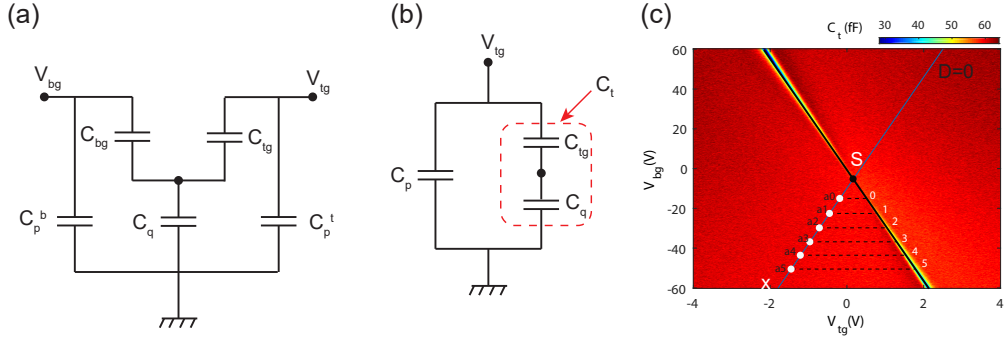

Fig. S4: (Color Online) (a) Schematic of the equivalent capacitive circuit of a dual gated BLG device. $C_p{}^b$ ($C_p{}^t$) are the parasitic capacitances. (b) Simplified equivalent circuit. $C_t$ is the measured total capacitance. (c) Extraction procedure to convert gate voltages ($V_{bg}$, $V_{bg}$) to Fermi energy ($E_F$) along constant $\bar{D}$ line.

## V. ESTIMATION OF $C_g$ FROM MAGNETO CAPACITANCE DATA

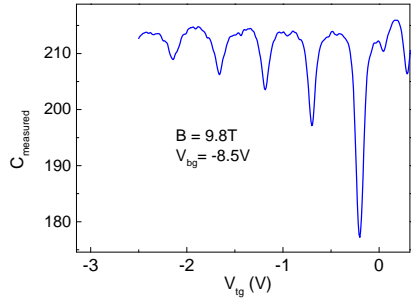

Fig. S5: (Color Online) Measured total capacitance ($C_t$) including parasitic capacitance as a function of $V_{tg}$ for $V_{bg} = -8.5$ $V$ at B=9.8T. The geometric capacitance was determined from the spacing between the capacitance minima, which is the charge required to fill each Landau level as discussed in the main text.

## VI. CONVERSION OF GATE VOLTAGES TO FERMI ENERGY IN FIG. 3A - FIG. 3B

In Fig. 3a of the main text, we present the magneto-capacitance data along the $D = 0$ line, where we sweep the gate voltages ($V_{tg}$, $V_{bg}$) synchronously along the $\bar{D} = 0$ line and measure the capacitance as a function of magnetic field. This means that $x-$ axis in Fig. 3a corresponds to specific values of ($V_{tg}$, $V_{bg}$). We know the capacitive coupling between the two gates as discussed in the main text, $C_{tg}/C_{bg} \sim 27$.

$$C_{tg}V_{tg'} = C_{bg}V_{bg} \tag{3}$$

implying $V_{tg'} = \frac{C_{bg}}{C_{tg}}V_{bg}$, therefore we convert $V_{bg}$ to $V_{tg'}$. Thus the total applied gate voltage can be written as $V_t = V_{tg} + V_{tg'}$. We now integrate the total capacitance ($C_t$) as a function of $V_t$ to extract the Fermi energy using the charge conservation relation $E_F = e \int_0^{V_t} \left(1 - \frac{C_t}{C_g}\right) dV_t$, which has been shown in Fig. 3b of the main text.


\end{document}